\documentstyle[12pt]{article}
\title{
Crossover from BCS to composite boson \\ (local pair) superconductivity
in quasi-2D systems}
\author{{\sl E.V.~Gorbar, V.M.~Loktev$^{*}$ and S.G.~Sharapov}\\
{\sl N.N.~Bogolyubov Institute for Theoretical Physics,} \\
{\sl 252143 Kiev-143, Ukraine}}
\date{}
\begin{document}
\maketitle

\begin{abstract}
The crossover from cooperative Cooper pairing to independent
bound state (composite bosons) formation and condensation in
 quasi-2D systems is studied.  It is shown that at low carrier
density the critical superconducting temperature is equal to the
temperature of Bose-condensation of ideal quasi-2D Bose-gas with
heavy dynamical mass, meanwhile at high densities the BCS result
remains valid. The evident nonmonotoneous behaviour of the  critical
temperature as function of the coupling constant (the energy of
the two particle bound state) is a qualitative difference of quasi-2D
crossover from 3D one.
\end{abstract}


\vfill

{\sl * Corresponding author: V.M. Loktev \\
     Bogolyubov Institute for Theoretical Physics \\
     Metrologicheskaya Str. 14-b \\
     252143, Kiev-143, Ukraine \\
     E-mail: eppaitp@gluk.apc.org}

\eject

	1. The problem of crossover from the BCS theory with cooperative Cooper
pairing to formation and condensation of composite bosons
\cite{history,Nozieres} (see also the review \cite{Micnas}) remains of interest
for a long time. After the discovery of high-temperature
superconductivity (HTSC) much attention was paid
to this problem \cite{Randeria2D,Gorbar,RanderiaBose}.
For example, in \cite{Gorbar},
using the Bethe-Salpiter equation, it was shown that
real bound states in {\bf r}-space are formed in a 2D-superconductor as
the fermion chemical potential $\mu$ takes negative values.

	For pure 2D superconductors, the transition from the regime of local
pairs to the Cooper pairing was studied in Refs. \cite{Randeria2D,Gorbar}
in the mean field (MF) approximation. However, it is well known that this
approximation  for a 2D problem is not  valid  for instance because of the
following reasons:

       $\imath)$ at first, by virtue of the Coleman-Mermin-Wagner-Hohenberg theorem
\cite{theorem} a "super-behaviour" is not possible for pure 2D systems
\cite{Kosterlitz} due to the fluctuations of order parameter (especially of
its phase) \cite{Rice}. Of course, any real system is not 2D  but at
least a quasi-2D one which, to be more specific, was neglected
in Refs. \cite{Randeria2D,Gorbar}, where the case of $T = 0$ was considered.

       $\imath \imath)$ at second, even for 3D systems the MF approximation in
the case of local pairs  superconductivity proves to be
wrong  \cite{RanderiaBose}: in fact, instead of the true
critical temperature $T_c$ it results in the value of the temperature
of local pair formation.

	Taking into account the Gauss fluctuations, a study of the problem of
crossover was carried out for a 3D system in Ref. \cite{RanderiaBose}. However,
it necessary to note that the HTSCs, study of such a crossover for which
is the most actual now, are not
ordinary 3D systems but  quasi-2D ones because of the essential
(up to five orders) anisotropy of
the conductivity. Therefore, from our point of view, the generalization of
results of Ref. \cite{RanderiaBose} for the case of quasi-2D systems would be
of significant interest.

        In this communication we study a model of a quasi-2D
superconductor taking into account the fluctuations of order parameter in
the random phase approximation using the Coleman-Weinberg method of effective
potential. We show that in the case of high carrier density
$n_{f}$ when $\mu \to \epsilon _{F}$ ($\epsilon _{F}$ is the Fermi
energy) the critical temperature $T_c$ is determined by the usual
BCS formula and  increases with the growth of attraction between
fermions. In the opposite limit -- the case of local pairs --
$T_c$ is determined by the formula for the temperature of
Bose-condensation of quasi-2D Bose-gas \cite{Wen}.
It is important that in the latter case $T_c$
decreases with the growth of such an attraction.
It is likely that the intermediate region which lies between
limiting cases abovementioned is most relevant to HTSCs for which
the value of $T_c$ is probably close to its maximal one.

	2. The simplest model Hamiltonian for carriers (electrons or holes)
in the system reads
\begin{equation}
H = -\psi_{\sigma}^{\dagger}(x) \left[ \frac{\nabla _{\perp}^{2}}{2 m_{\perp}} -
     \frac{1}{m_{z} d^{2}} \cos{(\imath \nabla _{z} d)} + \mu \right]
     \psi_{\sigma}(x) -
       V \psi_{\uparrow}^{\dagger}(x) \psi_{\downarrow}^{\dagger}(x)
       \psi_{\downarrow}(x) \psi_{\uparrow}(x),
                                                          \label{Hamilton}
\end{equation}
where $x \equiv t, \mbox{\bf r}_{\perp}, r_{z}$ (with $\mbox{\bf r}_{\perp}$
being a 2D vector); $\psi_{\sigma}(x)$ is a fermion field in the Shr\" odinger
representation, $\sigma$ is the spin variable;
$m_{\perp}$ is an effective mass of carriers in planes
(for example, CuO$_{2}$ ones);
$m_{z}$ is an effective mass in $z$-direction; $d$ is the interlayer distance;
$V$ is an effective local attraction inside planes;
$\mu$ and $n_{f}$ are determined above;
one also assumes that $\hbar = k_{B} = 1$.
The Hamiltonian proposed proves to be very convenient for study of
fluctuation stabilization by weak 3D one-particle inter-plane
tunneling. We omitted the two-particle tunneling considering it
smaller than the one taken into account.

        It is significant that the large anisotropy of conductivity cannot
be identified with the similar anisotropy of the effective masses.
In particular,
HTSCs with large anisotropy in $z$-direction do not display the metal
behaviour, which shows that interplane motion of particles is incoherent.
But, as will be seen, for justification of approximations used below the
not very large ratio $m_{z} / m_{\perp} \geq 10^{2}$ is already sufficient.
Namely such a value takes place, for instance, in HTSC 1-2-3.

	Lagrangian density connected with (\ref{Hamilton}) has the following
standard form:
\begin{displaymath}
L = \imath \psi_{\sigma}^{\dagger} \partial _{t} \psi_{\sigma} - H,
    \qquad \psi_{\sigma} \equiv \psi_{\sigma}(x).
\end{displaymath}

         Introducing an auxiliary Habbard-Stratonovich field
$\Phi \equiv V \psi_{\uparrow}^{\dagger} \psi_{\downarrow}^{\dagger}$ and
integrating over the fermions, one obtains the following
effective action
\begin{equation}
S_{eff}(\Phi, \Phi^{\ast}) = - \imath \mbox{TrLn}G^{-1} - \frac{1}{V}
                      \int dt d\mbox{\bf r} |\Phi|^{2}               \label{Seff}
\end{equation}
with
\begin{eqnarray}
G^{-1} = \imath \frac{\partial}{\partial t} I &+& \left(
         \frac{\nabla _{\perp}^{2}}{2 m_{\perp}} -
       \frac{1}{m_{z} d^{2}} \cos{(\imath \nabla _{z} d)} + \mu \right) \tau _{3} +
        \nonumber \\
  &+&  \frac{\tau _{1} + \imath \tau _{2}}{2} \Phi +
       \frac{\tau _{1} - \imath \tau _{2}}{2} \Phi ^{\ast},     \label{Green}
\end{eqnarray}
where $I$ is unity matrix, $\tau _{j}$ are the Pauli matrices. The
effective potential $U_{MF}  (\Phi, \Phi ^{\ast})$ which
describes the system with homogeneous order parameter in the  MF
approximation is determined from (\ref{Seff}) by the following relation:
\begin{displaymath}
S_{eff}(\Phi, \Phi^{\ast}) = \int dt d\mbox{\bf r} [
T_{kin} (\Phi, \Phi^{\ast}, \nabla \Phi, \nabla \Phi^{\ast}) -
U_{MF}  (\Phi, \Phi ^{\ast}) ],
\end{displaymath}
where $T_{kin}$ is a power series in derivatives
\footnote{For the case of a 2D system at $T = 0$, the first terms of this
series was found in Ref. \cite{Gorbar}.}.

	Applying the usual finite-temperature formalism \cite{Shrieffer},
the equation 
$\partial U_{MF}  / \partial \Phi \mid _{\Phi = \Phi ^{\ast} = 0} = 0$
result in the standard gap equation
\begin{equation}
\frac{1}{V} = \int \frac{d\mbox{\bf k}}{(2 \pi)^3} \frac{1}{2\xi (\mbox{\bf k})}
\tanh{\frac{\xi (\mbox{\bf k})}{2 T_{c}^{MF}}},                 \label{gap}
\end{equation}
where $\xi (\mbox{\bf k}) = \varepsilon (\mbox{\bf k}) - \mu$ with
$\varepsilon (\mbox{\bf k}) = \mbox{\bf k}_{\perp}^{2} /2 m_{\perp} + 1 /(m_{z} d^{2})
\cos{(k_{z} d)}$ (reason by which we use $T_{c}^{MF}$, rather than $T_c$
will be clarified in what follows).
We would like to emphasize that under temperatures of interest, the band width
in the $z$-direction is $m_{z}^{-1} d^{-2} \ll T_{c}^{MF}$.
As for the last inequality, it is easy to see, that at
$m_{z} \approx 10^{2} m_{e}$ and $d = 10 \dot A$ the value
$\hbar^{2} / (m_{z} d^{2} k_{B}) \sim 10$K is really more less
then the usual critical temperatures in HTSC compounds.

          To study Eq. (\ref{gap})
in the case of the local pairs superconductivity
the cutoff of BCS type can not be applied.
In order to eliminate the divergences in (\ref{gap})
it is convenient to introduce
(see Miyake in \cite{history} and \cite{Randeria2D,Gorbar})
the energy  of two-particle bound state
$\varepsilon _{b} = - 2 W \exp{(-4 \pi d / m_{\perp} V)}$, where $W =
 \mbox{\bf k}_{\perp max}^{2} /2 m_{\perp}$
is the band width in the plane
\footnote{
Note that in the region of parameters considered these bound states
form thresholdlessly.}.
Then, expressing the left-hand side of Eq. (\ref{gap}) through
$|\varepsilon _{b}|$ and passing to the limit  $W \to \infty$,
one can easy regularize the gap equation. The physical meaning of this
limiting procedure is based on enough natural assumption that
$W \gg \epsilon_{F}$.

	Contrary to the usual BCS approach in which it is always assumed that
$\mu = \epsilon _{F}$, here the value of chemical potential should be
consistently defined from the equation \\
$\partial U_{MF}  / \partial \mu \mid _{\Phi = \Phi ^{\ast} = 0} = -n_{f}$,
which leads to the second, or number, equation:
\begin{equation}
n_{F}(\mu , T_{c}^{MF}) \equiv \int \frac{d\mbox{\bf k}}{(2 \pi)^3}
     \left[ 1 - \tanh{\frac{\xi (\mbox{\bf k})}{2 T_{c}^{MF}}} \right]
                       = n_{f}.                         \label{number}
\end{equation}
In such a case one needs to solve simultaneously
the system of Eqs. (\ref{gap}), (\ref{number}) with two unknown
variables -- $T_{c}^{MF}$ and $\mu \equiv \mu (T_{c}^{MF})$, respectively
(in general case of arbitrary $T$ chemical potential $\mu \equiv \mu (T)
\neq \epsilon _{F}$).

        At high concentrations $n_{f}$, such that $\mu \gg
T_{c}^{MF}$, equality $\mu \simeq \epsilon _{F}$ indeed is the
solution of Eq. (\ref{number}), where $\epsilon _{F} = \pi n_{f} d/m_{\perp}$
is the Fermi energy of free 2D fermions with
$\varepsilon (\mbox{\bf k}) \sim \mbox{\bf k}^{2}$.  Taking into
account the regularization procedure, it follows from (\ref{gap})
that
\begin{equation}
T_{c}^{MF} \simeq \gamma \pi ^{-1}
\sqrt{2 |\varepsilon _{b}| \epsilon _{F}}    \label{TcMF}
\end{equation}
with $\gamma \simeq 1.781$.  This is just
well known BCS result for a 2D metal \cite{Randeria2D,Gorbar}.

        In the opposite case of small concentrations, such as
$-\mu \gg T_{c}^{MF}$, the roles of the gap (\ref{gap})
and number (\ref{number}) equations in some sense are reversed:
the Eq.  (\ref{gap}) determines $\mu$, while the Eq.
(\ref{number}) determines $T_{c}^{MF}$. Then, from the gap
equation, one can obtain that \\ 
$\mu = -|\varepsilon _{b}|/2 -
m_{z}^{-2} d^{-4} |\varepsilon _{b}|^{-1} / 2$.  
The latter except
for the second term corresponds to the result of Ref.
\cite{RanderiaBose}.  It is directly connected with the quasi-2D
character of the model and (despite this term is essentially less
than the first one) is very important in the case when the
fluctuations are taken into account.  At $|\varepsilon _{b}| \gg
\epsilon _{F}$ Eq. (\ref{gap}) transforms into the transcendent
one:  $|\varepsilon _{b}| / 2 T_{c}^{MF} = \ln{(T_{c}^{MF} /
\epsilon _{F})}$.  Herefrom, we directly obtain the abovementioned
result:  for a finite energy $\epsilon _{F}$, $T_{c}^{MF}$
strongly grows as  $V$ increases. Thus, for the case of small
carrier density the temperature $T_{c}^{MF}$ is not connected with
the critical one $T_c$ and in fact it corresponds to the
temperature of composite boson dissociation   \cite{RanderiaBose}.

	3. In order to take into account the effects of composite bosons formed
in the system, we apply the Coleman-Weinberg method of the
effective action \cite{Miransky}. In this method the effective potential
$U(\Phi , \Phi ^{\ast}) = U_{MF}  (\Phi, \Phi ^{\ast}) +
U^{(1)} (\Phi, \Phi ^{\ast})$,
where $U_{MF} $ is  the "tree-potential" (i.e., it is obtained in the MF
approximation) and $U^{(1)}$ is one-loop correction to it. At arbitrary $T$'s
\begin{eqnarray}
U^{(1)} (\Phi, \Phi ^{\ast}) = \frac{-\imath}{2}
\int \frac{d^{4} q}{(2 \pi)^4} \ln\left [ \left ( \frac{1}{V} -
\frac{\imath }{(2 \pi)^{4}} \int d^{4} p
\mbox{tr}[G(p) \tau_{+} G(p + q) \tau_{-}]
\right )^{2}  \right.
\nonumber                                                   \\
\left.   - \left ( \frac{\imath}{(2 \pi)^{4}}\int d^{4} p
\mbox{tr} [ G(p) \tau_{+} G(p + q) \tau_{+}] \right)^{2} \right ] ,
\qquad p \equiv p_{0}, \mbox{\bf p},
\quad q \equiv q_{0}, \mbox{\bf q},
                                                          \label{U(1)}
\end{eqnarray}
where
$p_{0} = \imath \pi (2 j +1) T$, $q_{0} = \imath 2 \pi l T$ ($j,
l$ -- integers), $\tau _{\pm} = (\tau _{1}  \pm \imath \tau _{2})
/ 2$, $G(p)$ and $G(p + q)$ are one-particle Green functions from
(\ref{Green}) in the frequency and momentum representation.
Expression (\ref{U(1)}) can be useful for the study the dependence
of energy gap $|\Phi |$ on the temperature.  In what follows, we
restrict ourselves  by the case of critical line $\Phi = \Phi
^{\star} = 0$ only and investigate the effect of the fluctuations
on $T_c$. In this case one obtains
\begin{displaymath}
U^{(1)}
(\Phi = \Phi ^{\ast} =0) = -\imath \int \frac{d^{4} q}{(2 \pi)^4}
\ln{D^{-1} \{ q_{0}, \mbox{\bf q} \} },
\end{displaymath}
where
\begin{eqnarray}
D^{-1} \{ q_{0}, \mbox{\bf q} \}  =  \frac{1}{V}
- \frac{1}{2} \int \frac{d \mbox{\bf p}}{(2 \pi)^{3}} \frac{1}
{\xi (\mbox{\bf p} - \mbox{\bf q} / 2) +
\xi (\mbox{\bf p} + \mbox{\bf q} / 2) - q_{0}}
\times
\nonumber                                                   \\
\left [ \tanh{\frac{\xi (\mbox{\bf p} - \mbox{\bf q} / 2)}{2 T_{c}}} +
\tanh{\frac{\xi (\mbox{\bf p} + \mbox{\bf q} / 2)}{2 T_{c}}} \right ].
                                                          \label{D}
\end{eqnarray}
Further, one has to remove the divergences from (\ref{D}) applying
the same procedure of regularization as in case of the gap Eq. (\ref{gap}).

According to \cite{Nozieres}, we introduce the phase $\delta (\omega , \mbox{\bf q})
 \equiv - \arg D^{-1} \{\omega + \imath 0, \mbox{\bf q} \}$ and take into
consideration the contribution of fluctuations through the number equation
(\ref{number}) (see also \cite{Varma})
\begin{eqnarray}
n_{F} (\mu, T_c) & +  & 2 n_{B} (\mu, T_c) = n_{f},
\nonumber                                                   \\
n_{B} (\mu, T_c) & \equiv & \frac{1}{2} \int \frac{d \mbox{\bf q}}{(2 \pi)^{3}}
\int _{- \infty}^{\infty} \frac{d \omega}{\pi} n_{B} (\omega)
\frac{\partial \delta (\omega , \mbox{\bf q})}{\partial \mu},
                                                       \label{numberBoson}
\end{eqnarray}
where $n_{B} (\omega) \equiv [\exp{(\omega / T) - 1}]^{-1}$ is the  Bose
distribution function. One can see from (\ref{numberBoson}) that the
system of fermions is divided into two coexisting and dynamically bounded
systems: fermi-particles, or unbounded fermions, and local pairs,
or bosons.  Therewith the corresponding concentrations of fermi-
and bose-particles depend on $T$ and are known only in mean.

	4. Consider the influence of $n_{B} (\mu , T_c)$
on the behaviour of the system as the carrier density changes.
Then we should consistently solve the system of Eqs. (\ref{gap})
and (\ref{numberBoson}). At high enough carrier density such a contribution,
$n_{B} (\mu , T_c)$ in (\ref{numberBoson}), is small, and we obtain that
$T_c \simeq T_{c}^{MF}$. It is worth to note the following: in more consistent
scheme one should take into consideration
the correction $\partial U^{(1)} / \partial \Phi $ in the gap equation
(\ref{gap}), and consequently the fluctuations would modify it.
However, because of the conditions $\mu = \epsilon _{F} \gg T_c$ and
$m_{z} d^ {2}  T_c \gg  1$, one can easily convinced that this
correction changes $T_c$ rather weakly \cite{Kats}.  It follows,
from (\ref{TcMF})
(the $T_{c}^{MF}$ coincides with the expression obtained in Ref.
\cite{Gorbar} for a pure 2D superconductor due to  the band
narrowness in the $k_z$-direction) that $T_c$ increases with  the
growth of couple constant $V$.

        At small concentrations, such that $|\mu| / T_c \gg 1$,
the contribution of bosons, contrary to the previous case, is dominant and
so one obtains from
(\ref{numberBoson}) that
\begin{equation}
n_{B} (\mu, T_c) \equiv  \int \frac{d \mbox{\bf q}}{(2 \pi)^{3}}
n_{B} \left( \frac{\mbox{\bf q}_{\perp}^{2}}{4 m_{\perp}} +
\frac{1}{2 |\varepsilon _{b}| (m_{z} d^{2})^{2}} [1 - \cos{(q_z d)}] \right )
\simeq \frac{n_{f}}{2},
                                                       \label{onlyBosons}
\end{equation}
i.e. all fermions are bounded between themselves forming composite bosons.
It is easy to see that the boson
effective mass  for its motion in the plane
retains the value $2 m_{\perp}$. As to the motion between the planes,
the effective boson mass
increases considerably: $2 |\varepsilon _{b}| m_{z}^{2} d^{2} ( \gg m_{z})$.
It is worth to stress that
this increasing has the dynamical character what is simply testified
by the presence of $|\varepsilon _{b}|$. Physically, it is ensured by the
one-particle character (see Eq. (\ref{Hamilton})) of the interplane tunneling.
Consequently, it is connected for a pair, with its virtual breakup, for which
the  energy  loss is of the order $|\varepsilon _{b}|$.

Now, one can use the formula for Bose condensation temperature of
ideal quasi-2D Bose-gas \cite{Wen}, it is easy to write down an
equation for determining $T_c$, which in the case  under
consideration takes the simple form
\begin{equation}
T_c \simeq \frac{\pi n d}{2 m_{\perp} \ln{(2 T_c |\varepsilon _{b}| m_{z}^{2} d^{4})}}
= \frac{\epsilon _{F}}{2 \ln{(2 T_c |\varepsilon _{b}| m_{z}^{2}
d^{4})}}.
\label{Tbose}
\end{equation}
The last equation
describes the characteristic properties of a quasi-2D
superconductor with  small carrier density: $\imath)$ at first, as the
critical temperature $T_c \sim \epsilon _{F}$, or (see above)
$T_c \sim n_{f}$, just as it should be  in 2D case (remind that
in a 3D one $T_c \sim n_{f}^{2/3}$ \cite{RanderiaBose}), and in MF
approximation 2D $T_{c}^{MF} \sim \sqrt{n_{f}} \gg T_{c}$
\cite{Gorbar}; $\imath \imath)$ secondly, contrary to the case of
3D superconductor where $T_c$ does not depend on $V$  at all
\cite{RanderiaBose}, in a quasi-2D system $T_c$ does depend on
$V$, namely: $T_c$ decreases with the growth of $V$.  As it was
stated above, the reason for this is the dynamical increasing of
the composite boson mass along the third direction. Thus, the growth of
$|\varepsilon _{b}|$ (or of $V$, what is the same) "makes" the system more
and more two-dimensional one even for the simplest case of quasi-2D metal
with a local four-fermion interaction.

It is interesting to note that decreasing $T_c$ can also take place in the case
when the local pairs are bipolarons \cite{Alexandrov};
then, the increasing of coupling with phonons, which makes the pairs more
massive, also leads to $T_c$ decreasing, (rather than increasing) as the
electron-phonon coupling grows.

	5. Our main results are not only expressions (\ref{TcMF}) and
(\ref{Tbose}) for $T_c$ in different limiting cases, namely, in
the cases of large and small $V$. No less important is comparison
of these expressions which show that for the given density of
fermions (the given $\epsilon_{F}$) there is two essential
regions.  If $|\varepsilon_{b}| \ll \epsilon_{F}$, then even in
the case of small (by absolute value $n_{f}$) densities the BCS
formula is valid and $T_c$ grows with the increasing of
$|\varepsilon_{b}|$ (see (\ref{TcMF})). In the opposite case,
$|\varepsilon_{b}| \gg \epsilon_{F}$, $T_c$ decreases
with the growth of $|\varepsilon_{b}|$. It show that in the case of
quasi-2D systems (it seems that the HTSCs belong to its case)
$T_{c}(|\varepsilon_{b}|)$ has a maximum, and, consequently, there
is a region of optimal (with respect to $|\varepsilon_{b}|$)
values of $\epsilon_{F}$ for which $T_c$ increases.  We may assume
that if we take into account the two-particle tunneling in
(\ref{Hamilton}) (if it does not exceed the one-particle
tunneling), then we obtain only the lower boundary for $T_c$ for
large $|\varepsilon_{b}|$. The region of
$\epsilon_{F} \approx |\varepsilon_{b}|$ needs a special
study because of both
presence of strongly developed fluctuations and  possible distinction of
properties of such a Fermi-liquid from the  Fermi-liquid of Landau type.
Note, in conclusion, that the fluctuations might also enhance non-monotonic
character of the $T_c$ behaviour versus $n_{f}$ obtained for
many-layers HTSC models \cite{GLShband}.

\section*{Acknowledgments}

        We are grateful to Profs.~V.P.~Gusynin, E.A.~Pashitskii,
I.I.~Ukrainskii
and Dr.~I.A.~Shovkovy  for numerous discussion and fruitful remarks.
The research describe in this publication was made possible in part by
Grant No K5O100.

\end{document}